\documentclass[12pt]{article}
\usepackage{epsf}
\title{ {\bf
The $Z\rightarrow l^+ l^-$ and $W\rightarrow \nu_{l} l^+$ decays in the 
noncommutative standard model}}
\author{\vspace{1cm}\\
        {\bf E. O. Iltan}
        \thanks{E-mail address:
        eiltan@heraklit.physics.metu.edu.tr}
 \\
        Physics Department, Middle East Technical University \\
        Ankara, Turkey\\}

\date{}

\begin{document}
\setlength{\baselineskip}{24pt}
\maketitle
\setlength{\baselineskip}{7mm}
\begin{abstract}
We study $Z\rightarrow l^+ l^-$ and $W\rightarrow \nu_{l} l^+$ decays
in the standard model  including the noncommutative effects. We observe
that these effects appear in the flavor dependent part of the decay widths 
of the processes under consideration and therefore, they are more effective 
for the heavy lepton decays.
\end{abstract} 
\thispagestyle{empty}
\newpage
\setcounter{page}{1}
%%%
%%%
\section{Introduction}
Leptonic Z-decays are among the most interesting lepton flavor conserving 
(LFC) and lepton flavor violating (LFV) interactions. The improved
experimental measurements at present stimulates the studies of these
interactions. With the Giga-Z option of the Tesla project, there is a 
possibility to increase Z bosons at resonance \cite{Hawkings}. The processes 
$Z\rightarrow l^- l^+$ with $l=e,\mu,\tau$ are among the LFC decays and they 
exist in the SM, even in the tree level. The experimental
predictions for the branching ratios ($BR$s) of these decays are 
\cite{PartData}   
\begin{eqnarray}
BR(Z\rightarrow e^+ e^-) &=& 3.366 \pm 0.0081\,\%  
\nonumber \, , \\
BR(Z\rightarrow \mu^+ \mu^-) &=& 3.367 \pm 0.013\,\% 
\nonumber \, , \\ 
BR(Z\rightarrow \tau^+ \tau^-) &=& 3.360 \pm 0.015 \,\% \, ,
\label{Expr1}
\end{eqnarray}
and the tree level SM predictions are  
\begin{eqnarray}
BR(Z\rightarrow e^+ e^-) &=& 3.331\,\%  
\nonumber \, , \\
BR(Z\rightarrow \mu^+ \mu^-) &=& 3.331\,\% 
\nonumber \, , \\ 
BR(Z\rightarrow \tau^+ \tau^-) &=& 3.328 \,\% \, .
\label{Expr2}
\end{eqnarray}
This shows that the tree level contribution of the SM plays the main role 
within the experimental uncertainities. In the literature, there are various 
experimental and theoretical studies \cite{Kamon}-\cite{IltanZllFC}. In 
\cite{Botz} a method to determine the weak electric dipole moment was 
developed. The vector and axial coupling constants, $v_f$ and $a_f$, in 
Z-decays have been measured at LEP \cite{LEP}. In \cite{Stiegler}, various 
additional types of interactions have been performed and a way to measure 
these contributions in the process $Z\rightarrow \tau^- \tau^+$ was 
described. \cite{IltanZllFC} is devoted to the possible new physics effects 
to the process $Z\rightarrow l^+ l^-$, in the general two Higgs doublet 
model. 

$W\rightarrow \nu_l l^+$ ($l=e,\mu,\tau$) decays exist also in the tree
level, in the SM and the experimental predictions for the branching  
ratios are \cite{PartData} 
\begin{eqnarray}
BR(W\rightarrow \nu_e e^+) &=& 10.9 \pm 0.4\,\%  
\nonumber \, , \\
BR(W\rightarrow \nu_{\mu} \mu^+) &=& 10.2 \pm 0.5\,\% 
\nonumber \, , \\ 
BR(W\rightarrow \nu_{\tau} \tau^+) &=& 11.3 \pm 0.8 \,\% \, ,
\label{Expr2W}
\end{eqnarray}
The main contribution to this decay comes from the SM in the tree level,
similar to the process $Z\rightarrow l^+ l^-$. There are large number of 
studies in the literature on this charged process \cite{Wnul}.

In the present work, we study $Z\rightarrow l^+ l^-$ and 
$W\rightarrow \nu_{l} l^+$ decays, with $l=e,\mu,\tau$, in the SM, 
including the noncommutative (NC) effects. The noncommutativity in the 
space-time is a possible candidate to describe the physics at very short 
distances of the order of the Planck length, since the nature of the 
space-time changes at these distances. In the noncommutative geometry the
space-time coordinates are replaced by Hermitian operators 
$\hat{x}_{\mu}$ which satisfy the equation \cite{Synder}
\begin{eqnarray}
[\hat{x}_{\mu},\hat{x}_{\nu}]=i\,\theta_{\mu\nu} \, ,
\label{com1}
\end{eqnarray}
where $\theta_{\mu\nu}$ is a real and antisymmetric tensor with the 
dimensions of length-squared. Here $\theta_{\mu\nu}$ can be treated as 
a background field and its components are assumed as constants over
cosmological scales.

It is possible to pass to the noncommutative field theory by introducing 
$*$ product of functions, instead of the ordinary one,
\begin{eqnarray}
(f*g)(x)=e^{\frac{i}{2}\,\theta_{\mu\nu} \,\partial^y_{\mu}\,
\partial^z_{\nu}} f(y)\, g(z)|_{y=z=x}\,.
\label{product}
\end{eqnarray}
The commutation of the Hermitian operators $\hat{x}_{\mu}$ (see eq.
(\ref{com1})) holds with this new product, namely, 
\begin{eqnarray}
[\hat{x}_{\mu},\hat{x}_{\nu}]_*=i\,\theta_{\mu\nu} \,\, .
\label{com2}
\end{eqnarray}

With the re-motivation due to the string theory arguments 
\cite{Connes2,Witten}, various studies on the noncommutative field 
theory (NCFT) have been done in the literature. However, NCFT have a
non-local structure and the Lorentz symmetry is explicitly violated. 
The violation of the Lorentz symmetry has been handled in 
\cite{Mocioiu,Carlson1} and  bounding noncommutative QCD due to the Lorentz 
violation has been studied in \cite{Carlson1}. In this work, it was
emphasised that the collider limits were not competitive with low energy 
tests of Lorentz violation for bounding the scale of space-time 
noncommutativity. Furthermore, the renormalizability and the unitarity  of
NC theories have been studied in the series of works \cite{Gonzales},
\cite{Gomis}, \cite{Hewett} and \cite{Chu}. The noncommutative quantum 
electrodynamics (NCQED) has been examined in \cite{Hayakawa,Riad} and the 
noncommutativity among extra dimensions for QED has been studied in 
\cite{Carlson2}. Furthermore, the noncommutativity in non-abelian case has
been formulated in \cite{Madore} and this formulation has been applied to 
the SM in \cite{Calmet}. Recently, a unique model for strong and electroweak
interactions with their unification has been constructed in \cite{Xiao}. In 
the work \cite{Behr}, the SM forbidden processes $Z\rightarrow \gamma\gamma$ 
and $Z\rightarrow gg$  has been studied by
including the NC effects. In \cite{Iltanbsgl}, the form factors, appearing 
in the inclusive $b\rightarrow s g$ decay, has been calculated in the NCSM, 
using the approximate phenomenology and the new operators existing in 
$b\rightarrow s g$ decay due to the NC effects has been obtained in
\cite{Iltanbsgam}. In the recent work, the possible effects of NC geometry 
on weak CP violation and the untarity triangles has been examined \cite{Cang}. 

The paper is organized as follows:
In Section 2, we present the explicit expressions for the branching ratios
of $Z\rightarrow l^+ l^-$ and $W\rightarrow \nu_l l^+$ in the framework of 
the NCSM. Section 3 is devoted to discussion and our conclusions.
%%%
%%%
\section{The noncommutative effects on the $Z\rightarrow l^+ l^-$ and 
$W \rightarrow \nu_l l^+$ decays in the SM}
The flavor conserving $Z\rightarrow l^+ l^-$, $l=e,\mu,\tau$, decays appear 
in the tree level in the SM. When the non-commutative effects are switched
on there exists a new contribution which is proportional to the a function
of the noncommutative parameter $\theta$. Our starting point is the effective 
action \cite{Calmet}
\begin{eqnarray}
S_{Matter, leptons}&=&\int \, d^4 x \Bigg( \sum_i 
(\bar{L}_L^{(i)}+\bar{L}_L^{(i) 1}+\bar{L}_L^{(i) 2})* i\, 
(\slash \!\!\!\! D^{SM}+\slash \!\!\!\! \Gamma) * (L_L^{(i)}+L_L^{(i) 1}+ 
L_L^{(i) 2}) 
\nonumber  \\
&+&\sum_i (\bar{e}_R^{(i)}+\bar{e}_R^{(i) 1}+\bar{e}_R^{(i) 2})* i\, 
(\slash \!\!\!\! D^{SM}+\slash \!\!\!\!\Gamma) * (e_R^{(i)}+e_R^{(i) 1}+ 
e_R^{(i)2}) \Bigg)+ O(\theta^3)\, ,  
\label{S1}
\end{eqnarray}
with
\begin{eqnarray}
D_{\mu}^{SM} L_L &=& 
(\partial_{\mu}-i g' Y_L A_{\mu}- i g B_{\mu a} T^a_L) L_L 
\nonumber \, , \\
D_{\mu}^{SM} e_R &=& 
(\partial_{\mu}-i g' Y_R A_{\mu}) e_R \, ,
\label{DSM}
\end{eqnarray}
and 
\begin{eqnarray}
L_L^{(i) 1}&=&-\frac{1}{2} \theta_{\mu\nu} 
(g' Y_L A^{\mu}+g B^{\mu}_ a  T^a_L) \, \partial^{\nu} L_L^{(i)}+ 
O(A^2,B^2,A\,B) 
\nonumber \,, \\
L_L^{(i) 2}&=&-\frac{i}{8} \theta_{\mu\nu}\, \theta_{\alpha\beta}\, 
( g' Y_L \, \partial^{\mu} A^{\alpha} + g\, \partial^{\mu} B^{\alpha}_ a 
T^a_L)\, 
\partial^{\nu}\,\partial^{\beta} L_L^{(i)}+ O(A^2,B^2,A\,B) 
\, , \nonumber \\ 
e_R^{(i) 1}&=&-\frac{1}{2} \theta_{\mu\nu}\, (g' Y_R A^{\mu}) \,
\partial^{\nu} e_R^{(i)}+ O(A^2) 
\nonumber \,, \\
e_R^{(i) 2}&=&-\frac{i}{8} \theta_{\mu\nu}\, \theta_{\alpha\beta} \, 
(g' Y_R \partial^{\mu} A^{\alpha})\, 
\partial^{\nu}\,\partial^{\beta} e_R^{(i)}+  O(A^2) \, ,
\label{L12eR12}
\end{eqnarray}
where $*$ in eq. (\ref{S1}) denotes the Moyal-Weyl star product (see eq.
(\ref{product})), $L_L^{(i)}$ ($e_R^{(i)}$) is the left (right) handed lepton 
doublet of $i^{th}$ family, $Y_L=-\frac{1}{2}$, $Y_R=-1$ and 
$O(A^2,B^2,A\,B)\, ( O(A^2))$ 
is the part of $L_L^{(i) 1,2}$ ($e_R^{(i) 1,2}$) which includes the 
interactions of more than one gauge fields. Here the function  $\Gamma$ has 
no interest since it contains two gauge field interactions, which do not 
give any contribution to our processes  
$Z (W) \rightarrow l^+ l^- (\nu_l l^+)$. Furthermore, we do not present 
the parts of $L_L^{(i) 1,2}$ and $e_R^{(i) 1,2}$, $O(A^2,B^2,A\,B)$ and 
$O(A^2)$, which include the interactions of more than one gauge fields 
(see \cite{Madore} and \cite{Calmet} for details).     

Finally, the additional vertex to the $Z\rightarrow l^+ l^-$ decay to the 
second order in $\theta$ can be obtained as  
\begin{eqnarray}
V_{\mu, NC}^Z&=& \Bigg( 
(\theta_{\mu\nu} \gamma_{\alpha}+\theta_{\nu\alpha} \gamma_{\mu}+ 
\theta_{\alpha\mu} \gamma_{\nu})\, p_Z^{\nu} p_1^{\alpha}-
\frac{i}{4}\,(\theta_{\mu\nu} \gamma_{\alpha}+
\theta_{\nu\alpha} \gamma_{\mu}+ \theta_{\alpha\mu} \gamma_{\nu})\, 
\theta_{\gamma\sigma}\, p_Z^{\gamma}\, p_Z^{\alpha}\, p_1^{\sigma}\, 
p_1^{\nu} \Bigg ) \times \nonumber \\ & &  (c_1\, L+c_2\,R) \, ,
\label{vertexZllNC}
\end{eqnarray}
where $c_1=-e \frac{2 sin^2\theta_W-1}{4\, sin \theta_W cos\theta_W}$, 
$c_2=-e \frac{tan\theta_W}{2}$, $L(R)=\frac{1-\gamma_5}{2}\, 
(\frac{1+\gamma_5}{2})$ and $p_Z$ ($-p_1$) incoming (ougoing) four momentum 
of Z boson with polarization vector $\epsilon^{\mu}$ (anti-lepton). Notice 
that the part of the vertex proportional with $\theta_{\nu\alpha}$ would be 
the whole contribution in the case that the NC effects enter into the 
expressions as an exponential factor $e^{\frac{i}{2}\, \theta_{\mu\nu} 
p_Z^{\mu} p_1^{\nu}}$, which is consistent in approximate phenomenology 
(see \cite{Hinchliffe} and references therein). 

Now we present the BR of the process $Z\rightarrow l^+ l^-$ including the
non commutative effects at the least order in $\theta$, in the Z boson rest 
frame:
\begin{eqnarray}
BR=\frac{\alpha_{em}\,m_Z }{6\, \Gamma_Z\,sin^2\,2\theta_W} \Bigg ( 
(1-4\,sin^2\theta_W+\sin^4\theta_W )-\frac{m_l^2}{m_Z^2} 
(1+8\,sin^2\,\theta_W-16\,sin^4\,\theta_W+\frac{m_Z^4}{16}\, f(\theta))\Bigg) 
\, ,
\label{BRZll}
\end{eqnarray}
where $\Gamma_Z$ is the total decay width of Z boson, $\Gamma_Z=2.490\,
GeV$, and $\alpha_{em}=\frac{e^2}{4\, \pi}$. As shown in this equation, the NC effects
appear as the function of $\theta$, 
\begin{eqnarray}
f(\theta)=(\vec{\theta}_T.\hat{p}_1)^2+(\vec{\theta}_S.\hat{p}_1)^2
-(|\vec{\theta}_T|^2+|\vec{\theta}_S|^2)+2\,\hat{p}_1.
(\vec{\theta}_T \times \vec{\theta}_S) \, . 
\label{ftheta}  
\end{eqnarray}
Here we use the definitions $(\theta_T)_i=\theta_{0i}$ and $(\theta_S)_i= 
\frac{1}{2}\epsilon_{ijk} \theta^{jk}$, $i,j,k=1,2,3$ and
$\vec{p}_1=\frac{m_Z}{2}\hat{p}_1$. $(\theta_T)_i$ and $(\theta_S)_i$
are  responsible for time-space and space-space noncommutativity,
respectively. The noncommutative effects enter into expression with lepton 
mass and their effects are much more suppressed in the case of light leptons.  
Notice that, the terms of the vertex eq. (\ref{vertexZllNC}) which is second 
order in $\theta$ do not give any contribution to the $BR$  of the decay 
$Z\rightarrow l^+ l^-$, in the $Z$ boson rest frame. 

The charged $W\rightarrow \nu_l l^+$  decays exist with the charged current 
and they also appear at the tree level in the SM. Similar to the 
$Z\rightarrow l^+ l^-$ decay, the noncommutative effects are controlled by 
the additional vertex 
\begin{eqnarray}
V_{\mu, NC}^W &=&-\frac{e}{2\,\sqrt{2}\,sin \theta_W} \times \nonumber \\
& & \!\!\!\!\!\!\!\!\!\!\!\!\!\!\!\!\!\! \Bigg((\theta_{\mu\nu} \gamma_{\alpha}+\theta_{\nu\alpha} 
\gamma_{\mu}+ \theta_{\alpha\mu} \gamma_{\nu})\, p_W^{\nu} p_1^{\alpha}
-\frac{i}{4}\,(\theta_{\mu\nu} \gamma_{\alpha}+
\theta_{\nu\alpha} \gamma_{\mu}+ \theta_{\alpha\mu} \gamma_{\nu})\, 
\theta_{\gamma\sigma}\, p_W^{\gamma}\, p_W^{\alpha}\, p_1^{\sigma}
\, p_1^{\nu} \Bigg ) \,L \, ,
\label{vertexWnulNC}
\end{eqnarray}
where $p_W$ ($-p_1$) incoming (ougoing) four momentum of W boson with 
polarization vector $\epsilon^{\mu}$ (anti-lepton). The BR of the process 
$W\rightarrow \nu_l l^-$ including the non commutative effects, at the least 
order in $\theta$, in the W boson rest frame reads as:
\begin{eqnarray}
BR=\frac{\alpha_{em}\,m_W }{384\, \Gamma_W\,sin^2\,\theta_W} \Bigg ( 
(32+\frac{m_l^2}{m_W^2} (16-m_W^4\, f(\theta)) \Bigg) \, , 
\label{BRWnul}
\end{eqnarray}
where $\Gamma_W$ is the total decay width of W boson, $\Gamma_W=2.060\, GeV$. 
Here the function $f(\theta)$ (see eq. (\ref{ftheta})) represents the 
noncommutative effects. The terms of the vertex eq. (\ref{vertexWnulNC}) 
which is second order in $\theta$ give a non-zero contribution to the $BR$  
of the decay $W\rightarrow \nu_l l^+$, in the $W$ boson rest frame. This
contribution is proportional to 
$m_l^2\, m_W^2\, (\vec{\theta}_T.\hat{p}_1)^2$. However, it is cancelled by
the part, coming from the vertex linear in $\theta$.   

At this stage we try to parametrize the vectors $(\theta_T)_i$ and 
$(\theta_S)_i$ which are responsible for time-space and space-space 
noncommutativity, respectively. With the assumption that the matrix 
$\theta_{\mu\nu}$ is real and constant, we take 
\begin{eqnarray}
\vec{\theta}_T&=&A_1 \hat{p}_1+A_2 \hat{p}^T_{1\bot} \, , 
\\ \nonumber    
\vec{\theta}_S&=&B_1 \hat{p}_1+B_2 \hat{p}^S_{1\bot}\, , 
\label{thetaTS}  
\end{eqnarray}
where $\hat{p}_1$ ($\hat{p}^T_{1\bot}$, $\hat{p}^S_{1\bot}$) is the unit 
vector in the direction of (the perpendicular direction to) the incoming 
lepton three momentum $\vec{p}_1$ (for $\vec{\theta}_T$, $\vec{\theta}_S$), 
$A_i, B_i$ are the corresponding real coefficients. Using this 
parametrization, $f(\theta)$ can written as 
\begin{eqnarray}
f(\theta)= 2\, A_2\, B_2\,  \hat{p}_1. (\hat{p}^T_{1\bot}\times 
\hat{p}^S_{1\bot})-(A_2^2+B_2^2) \, .
\label{fthetanew}  
\end{eqnarray}
and this shows that the transverse components of the vectors 
$\hat{p}^T_{1\bot}$ and $\hat{p}^S_{1\bot}$ to the incoming lepton three 
momentum $\vec{p}_1$ play the main role for the NC effects. In the case
of $\hat{p}^T_{1\bot}\bot\hat{p}^S_{1\bot}\bot \hat{p_1}$ with $A_2=B_2$, 
the noncommutative effects are switched off. Furthermore, for 
$\vec{\theta}_T\parallel \hat{p}_1$ ($\vec{\theta}_S\parallel \hat{p}_1$),
the coefficient $A_2=0$ ($B_2=0$) and therefore, only the space-space 
(space-time) noncommutativity is responsible for the noncommutative effects. 
This is interesting in the determination of the noncommutative directions 
with the help of the future sensitive experimental results.      
%%%
%%%
\section{Discussion}
In this section, we analyse the NC effects on the $BR$ of the flavor conserving 
$Z\rightarrow l^+ l^-$ and charged $W\rightarrow \nu l^+$  decays, in the 
framework of the SM. The processes underconsideration exist in the tree level 
in the SM and the theoretical calculation of the BRs obey the experimental 
results within the measurement errors. 

The flavor $l=e,\mu,\tau$ dependence of the part of the $BR(Z\rightarrow
l^+ l^-)$ is extremely weak 
\begin{eqnarray}
R_{\mu e}&=& \frac{BR (Z\rightarrow\mu^+ \mu^-)}{BR (Z\rightarrow e^+ e^-)}
=1.0008 \pm 0.005\, , \nonumber \\
R_{\tau e}&=&\frac{BR(Z\rightarrow\tau^+ \tau^-)}{BR (Z\rightarrow e^+ e^-)}
=0.998\pm 0.005\, . 
\label{BRZllex2}
\end{eqnarray}
This part, which controls the flavor effects, is proportional to the factor 
$\frac{m_l^2}{m_Z^2}$ and it includes the noncommutative effects. Therefore, 
it is more informative to study the heavy lepton decays to determine the 
noncommutativity of the geometry. Notice that we choose the non-commutative
parameter $\theta=|\theta_{\mu\nu}|$ as at the order of the magnitude of 
$\sim 10^{-6}-10^{-5} \, GeV^{-2}$.

\begin{figure}[htb]
\vskip -3.0truein
\centering
\epsfxsize=6.8in
\leavevmode\epsffile{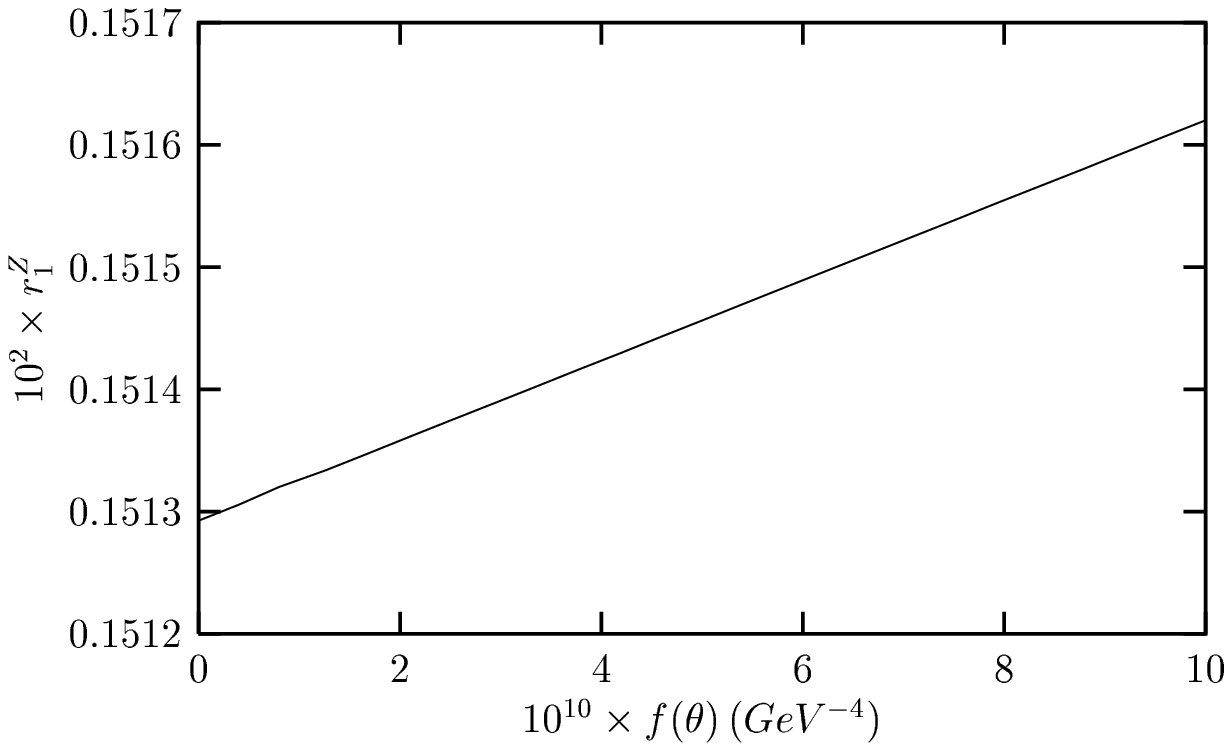}
\vskip -3.0truein
\caption[]{ $f(\theta)$ dependence of ratio $r_1^Z=\frac{BR_{flavor}}
{BR_{tot}}$ where $BR_{flavor}$ is the flavor dependent part of the 
$BR$ and $BR_{tot}$ is the total $BR$, for the process 
$Z\rightarrow \tau^+ \tau^-$.} 
\label{Zllratiomlasiltau}
\end{figure}

\begin{figure}[htb]
\vskip -3.0truein
\centering
\epsfxsize=6.8in
\leavevmode\epsffile{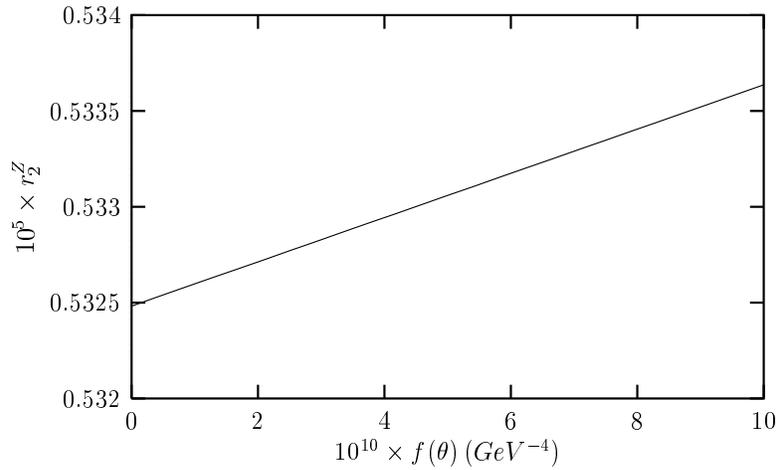}
\vskip -3.0truein
\caption[]{The same as Fig. \ref{Zllratiomlasiltau}
but for $Z\rightarrow \mu^+ \mu^-$ decay.}
\label{Zllratiomlasilmu}
\end{figure}

In Fig. \ref{Zllratiomlasiltau}, we present the noncommutative parameter 
$f(\theta)$ dependence of ratio $r_1^Z=\frac{BR_{flavor}}{BR_{tot}}$
where $BR_{flavor}$ is the flavor dependent part of the $BR$ and $BR_{tot}$ 
is the total $BR$, for the process $Z\rightarrow \tau^+ \tau^-$. This figure
shows that the noncommutative effects are at most at the order of the
magnitude of $0.001 \%$, even for the heavy lepton $\tau$ decay. This
dependence becomes extremely small, $10^{-6}\, \%$, for 
$Z\rightarrow \mu^+ \mu^-$ decay (see Fig. \ref{Zllratiomlasilmu} ), since 
the mass of the lepton $\mu$ is small and there is a strong suppression 
factor $\frac{m^2_{\mu}}{m^2_Z}$ for $BR_{flavor}$.  

\begin{figure}[htb]
\vskip -3.0truein
\centering
\epsfxsize=6.8in
\leavevmode\epsffile{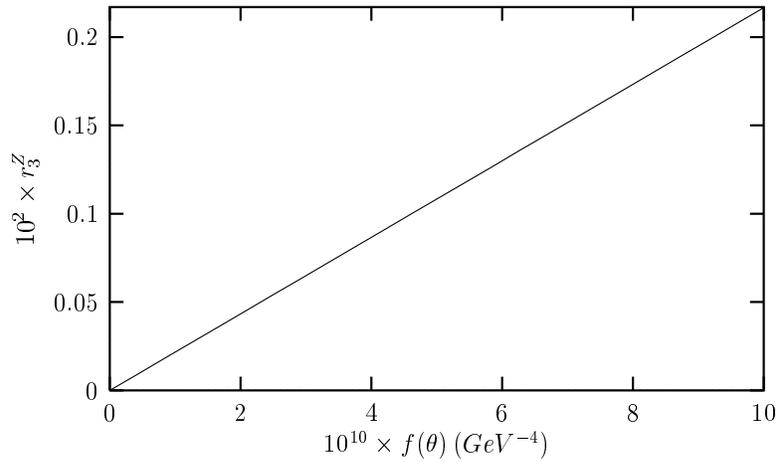}
\vskip -3.0truein
\caption[]{ $f(\theta)$ dependence of ratio $r_3^Z=
\frac{BR_{flavor_{\theta}}}{BR_{flavor}}$ where $BR_{flavor_{\theta}}$ is 
the noncummutative-flavor dependent part of the $BR$, for the process 
$Z\rightarrow \tau^+ \tau^-$.}
\label{Zllratiomltetmlasil}
\end{figure}

Fig. \ref{Zllratiomltetmlasil} is devoted to the $f(\theta)$ dependence of 
ratio $r_3^Z=\frac{BR_{flavor_{\theta}}}{BR_{flavor}}$ where 
$BR_{flavor_{\theta}}$ is the noncommutative-flavor dependent part of the 
$BR$, for the process $Z\rightarrow \tau^+ \tau^-$. It is observed that 
the noncommutative effects on the flavor dependent part can reach to 
$0.1 \%$. 

Now we would like to study the charged $W\rightarrow \nu_l l^+$ decay and
the noncommutative effects on this process. The $BR$ for this process is 
\begin{eqnarray}
BR(W\rightarrow \nu_l l^+)=10.74\pm 0.33\, \% 
\label{BRWnulex}
\end{eqnarray}
and the flavor $l=e,\mu,\tau$ dependence of this value is weak. 
Similar to the $Z\rightarrow l^+ l^-$ decay,  the part of the 
BR($W\rightarrow \nu l^+ $) which controls the flavor effects is 
proportional to the factor $\frac{m_l^2}{m_W^2}$ and the noncommutative 
effects appear in this part. 

\begin{figure}[htb]
\vskip -3.0truein
\centering
\epsfxsize=6.8in
\leavevmode\epsffile{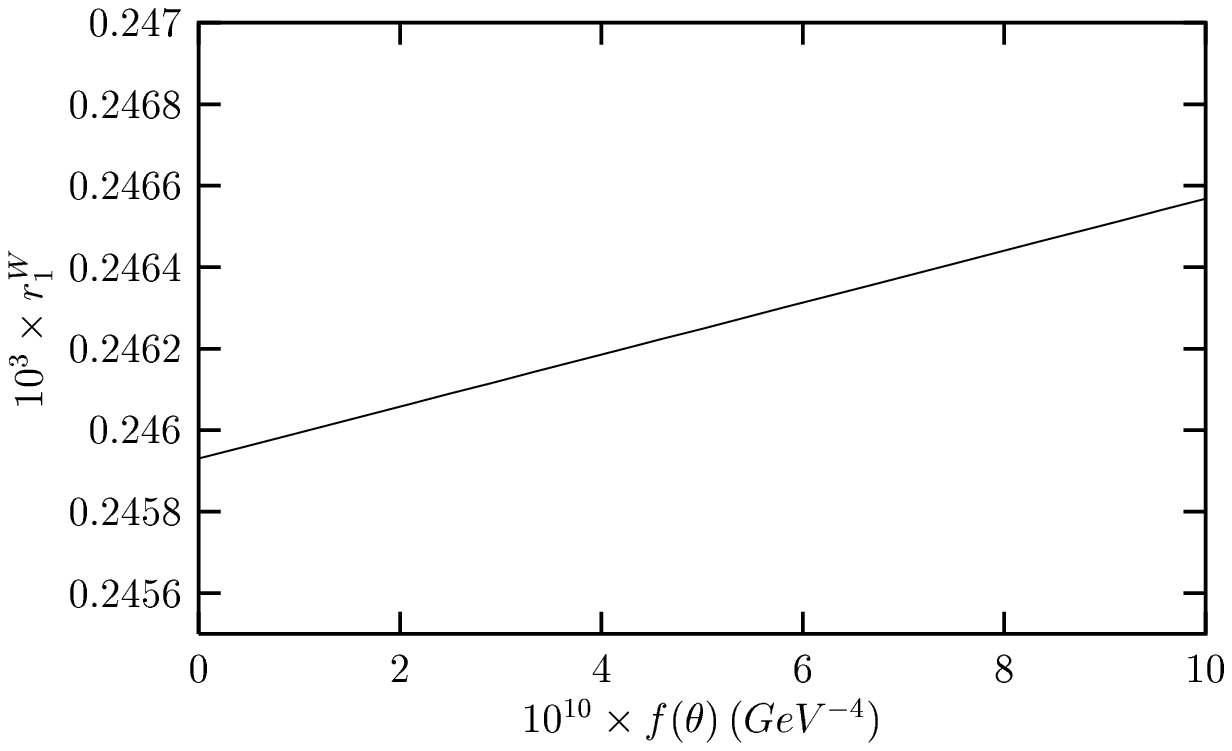}
\vskip -3.0truein
\caption[]{The same as Fig. \ref{Zllratiomlasiltau}
but for $W\rightarrow \nu_{\tau} \tau^+$ decay.}
\label{Wnulratiomlasiltau}
\end{figure}

In Fig. \ref{Wnulratiomlasiltau}, we present the noncommutative parameter 
$f(\theta)$ dependence of ratio $r_1^W=\frac{BR_{flavor}}{BR_{tot}}$
where $BR_{flavor}$ is the flavor dependent part of the $BR$ and $BR_{tot}$ 
is the total $BR$ for the process $W\rightarrow \nu_{\tau} \tau^+$. It is 
observed that the noncommutative effects are at most at the order of the
magnitude of $0.001 \%$, for the heavy lepton $\tau$ decay. 

\begin{figure}[htb]
\vskip -3.0truein
\centering
\epsfxsize=6.8in
\leavevmode\epsffile{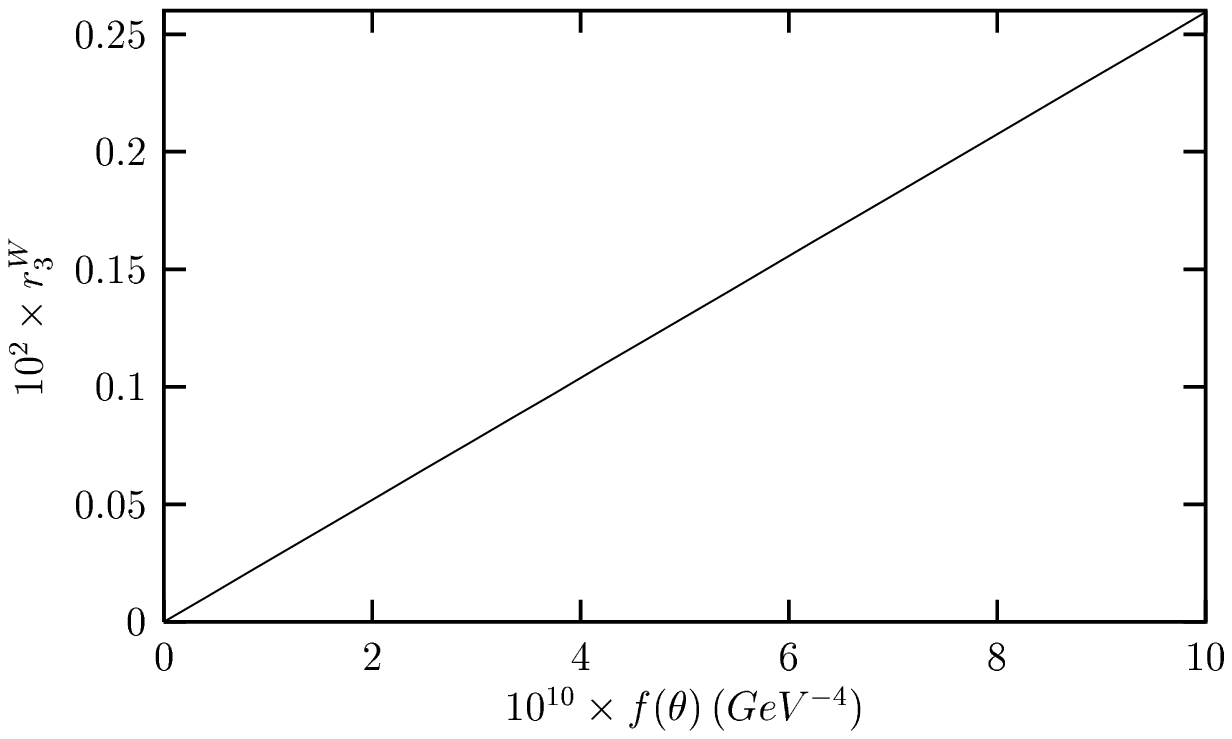}
\vskip -3.0truein
\caption[]{The same as Fig. \ref{Zllratiomltetmlasil}
but for $W\rightarrow \nu_{\tau} \tau^+$ decay.}
\label{Wnulratiomltetmlasil}
\end{figure}

Fig.\ref{Wnulratiomltetmlasil} represents the $f(\theta)$ dependence of 
ratio $r_3^W=\frac{BR_{flavor_{\theta}}}{BR_{flavor}}$ where 
$BR_{flavor_{\theta}}$ is the noncommutative-flavor dependent part of the 
$BR$ for the process $W\rightarrow \nu_{\tau} \tau^+$. Here, the 
noncommutative effects on the flavor dependent part can reach to 
$0.1 \%$, similar to the process $Z\rightarrow \tau^- \tau^+$. 

In conclusion, the NC effects in the decays under consideration are
effective in the flavor dependent part of their BRs. With the possible
future experiments, which are sensitive to the flavor dependent part of
these processes, those effects can be extracted and the noncommutative 
direction can be determined. 
%%%
%%%
\section{Acknowledgement}
This work was supported by Turkish Academy of Sciences (TUBA/GEBIP).
%%%
%%%
%

\newpage
\end{document}